%% file: main.tex
\documentclass[preprint,journal]{vgtc}       % preprint (journal style)

\ifpdf%                                % if we use pdflatex
  \pdfoutput=1\relax                   % create PDFs from pdfLaTeX
  \usepackage{graphicx}                % allow us to embed graphics files
  \DeclareGraphicsExtensions{.pdf,.png,.jpg,.jpeg} % for pdflatex we expect .pdf, .png, or .jpg files
\else%                                 % else we use pure latex
  \usepackage{graphicx}                % allow us to embed graphics files
  \DeclareGraphicsExtensions{.eps}     % for pure latex we expect eps files
\fi%

\graphicspath{{figures/}{pictures/}{images/}{./}} % where to search for the images

\usepackage{microtype}                 % use micro-typography (slightly more compact, better to read)
\PassOptionsToPackage{warn}{textcomp}  % to address font issues with \textrightarrow
\usepackage{textcomp}                  % use better special symbols
\usepackage{mathptmx}                  % use matching math font
\usepackage{times}                     % we use Times as the main font
\usepackage{cite}                      % needed to automatically sort the references
\usepackage{tabu}                      % only used for the table example
\usepackage{booktabs}                  % only used for the table example
\usepackage[normalem]{ulem}
\usepackage{amsmath}
\usepackage{amsfonts}
\usepackage{enumitem}
\usepackage{wrapfig}
\usepackage{pdfpages}
\usepackage{threeparttable,caption,tabularx,booktabs}
\newcolumntype{L}{>{\hsize=1.5\hsize\raggedright\arraybackslash}X}
\newcolumntype{R}{>{\hsize=0.9\hsize\raggedleft\arraybackslash}X}

\onlineid{0}

\vgtccategory{Research}
\vgtcpapertype{evaluation}

\title{Visual Reasoning Strategies\\ for Effect Size Judgments and Decisions}
\author{Alex Kale, Matthew Kay, and Jessica Hullman}
\authorfooter{
\item
Alex Kale is with the University of Washington. E-mail: kalea@uw.edu.
\item
Matthew Kay is with the University of Michigan. E-mail: mjskay@umich.edu.
\item
Jessica Hullman is with Northwestern University. E-mail: jhullman@northwestern.edu.
}

\shortauthortitle{Kale \MakeLowercase{\textit{et al.}}: Visual Reasoning Strategies for Effect Size Judgments and Decisions}
\abstract{
Uncertainty visualizations often emphasize point estimates to support magnitude estimates or decisions through visual comparison.
However, when design choices emphasize means, users may overlook uncertainty information and misinterpret visual distance as a proxy for effect size.
We present findings from a mixed design experiment on Mechanical Turk which tests eight uncertainty visualization designs: 95\% containment intervals, hypothetical outcome plots, densities, and quantile dotplots, each with and without means added. 
We find that adding means to uncertainty visualizations has small biasing effects on both magnitude estimation and decision-making, consistent with discounting uncertainty.
We also see that visualization designs that support the least biased effect size estimation do not support the best decision-making, suggesting that a chart user's sense of effect size may not necessarily be identical when they use the same information for different tasks.
In a qualitative analysis of users' strategy descriptions, we find that many users switch strategies and do not employ an optimal strategy when one exists.
Uncertainty visualizations which are optimally designed in theory may not be the most effective in practice because of the ways that users satisfice with heuristics, suggesting opportunities to better understand visualization effectiveness by modeling sets of potential strategies.
} % end of abstract


\CCScatlist{ % not used in journal version
 \CCScat{H.m}{Information Systems}%
{Miscellaneous}
}

\teaser{
  \centering
  \includegraphics[width=\linewidth]{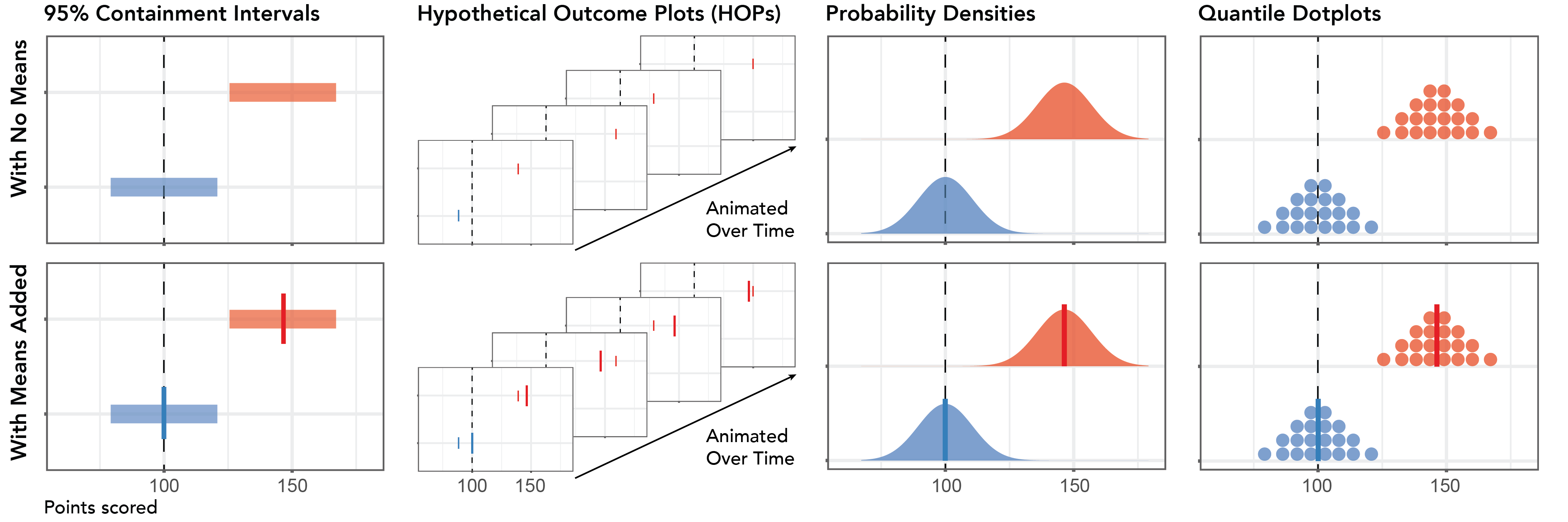}
  \setlength{\abovecaptionskip}{-10pt}
  \caption{Visualization designs evaluated in our experiment.
  } 
  \label{fig:teaser}
}

\vgtcinsertpkg

\begin{document}


\firstsection{Introduction}

\maketitle

\input{1_intro.tex}
\input{2_background.tex}

\input{3_method.tex}
\input{4_results.tex}
\input{5_strategies}
\input{6_discussion}

\acknowledgments{
We thank the members of the UW IDL and Vis-Cog Lab, as well as the MU Collective at Northwestern for their feedback.
This work was supported by a grant from the Department of the Navy (N17A-T004).}

\bibliographystyle{abbrv-doi}

\bibliography{7_es-jdm}
\end{document}

%% file: 1_intro.tex
%% \section{Introduction} %for journal use above \firstsection{..} instead
Many visualization authors perceive visualizing uncertainty as an exception, rather than a norm~\cite{Hullman2020}. 
However, the common practice of omitting uncertainty information from visualizations and focusing attention on point estimates leads to ``incredible certitude''~\cite{Manski2018a,manski2019lure}, the unwarranted impression that error is minimal or not important. 
To enable informed judgments and decisions, a common suggestion is to present uncertainty information alongside point estimates, for example, by showing intervals in which estimates could fall~\cite{Cumming2005,Cumming2014,Manski2018b,taylor_guidelines_1994}.

However, presenting uncertainty alongside point estimates may not lead users to incorporate uncertainty information into their judgments. 
A large body of work on biases due to heuristics (e.g.,~\cite{Tversky1974,Tversky1981,Kahneman2011}), also commonly known as \textit{satisficing}~\cite{Simon1956}, shows that people often avoid or discount uncertainty information.
% \jessica{suggestion - now say make clear that its possible to ignore uncertainty in favor of means when they are graphically presented. Then, jump to 'Different visualization design choices' paragraph. After that, then introduce distance heuristics, e.g But how might users of visualizations who focus on means judge effect size? Imagine a user viewing a visualization like 'example fig'. Discounting uncertainty may manifest as ... Then do the 'As with any' paragraph. Reason for this - you're jumping in to distance heuristic before setting up the necessary point about how someone can still ignore uncertainty when means are visually distinguishable in an uncertainty vis. it's too quick of a jump and I think readers will be confused.  }
This suggests that chart users may ignore uncertainty in favor of means even when both are presented~\cite{Hullman2015}.
% \idl{The intro really jumps right into the experimental design, talking about your hypothesis and charts you used. This may be okay, but it's slightly lower level than I would expect.} \alex{We may have addressed this by incorporating the edits that Jessica suggested, but we also have some space to extend this narrative if that would help.}

Different visualization design choices make the mean more or less salient. Imagine a continuum of uncertainty visualization designs representing how perceptually difficult it is to decode the mean from a chart. 
At one extreme are hypothetical outcome plots or HOPs~\cite{Hullman2015,Kale2019-hops} where the mean is only encoded implicitly as the average of a set of outcomes presented across frames of an animation.
At the other extreme are direct encodings of point estimates presented alongside uncertainty (e.g., represented as error bars). 
We expect that the salience of the mean in uncertainty visualization designs and other factors such as frequency-framing of probability~\cite{Hullman2015,Kay2016,Kale2019-hops,Fernandes2018} influence the degree to which users focus on means and ignore uncertainty.

\begin{figure}[t]
    \centering
    \includegraphics[width=\columnwidth]{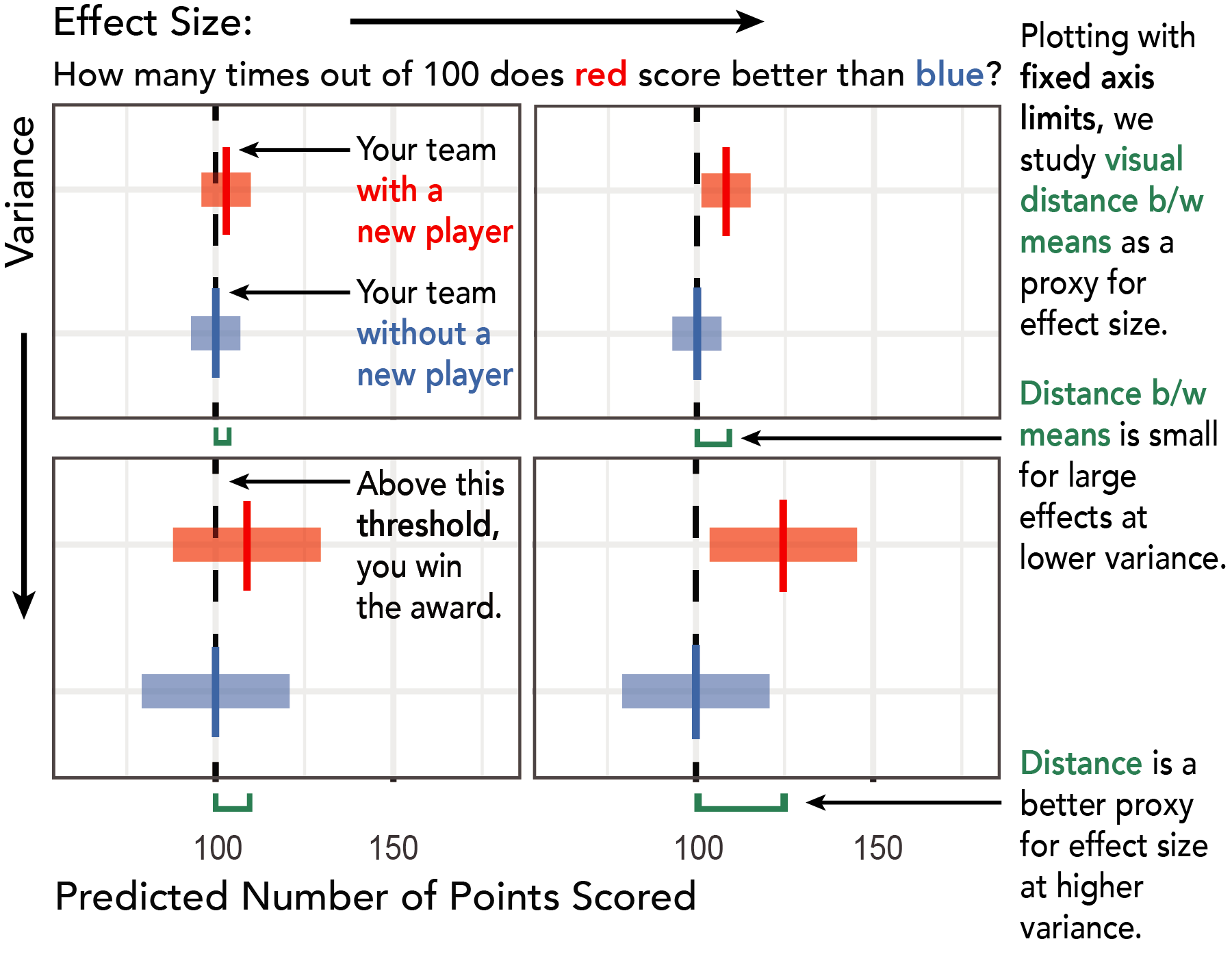}
    \setlength{\abovecaptionskip}{-10pt}
    \setlength{\belowcaptionskip}{-10pt}
    \caption{Intervals with means showing two levels of effect size (72\% and 95\% $\Pr(S)$) at low and high variance. Using visual distance between means as a proxy for effect size should result in greater bias toward underestimating effect size at lower variance than at higher variance.
    }
    \label{fig:distance_strategy}
\end{figure}

How might chart users who focus on means judge effect size? 
Imagine a user viewing visualizations like those in Figure~\ref{fig:teaser}. 
Discounting uncertainty may manifest as using \textit{distance between means or gist estimates of distance between distributions as a proxy for effect size} and not judging distance relative to the width of distributions.
Using only distance as a proxy for effect size may be misleading (Fig.~\ref{fig:distance_strategy}) because the distance between distributions depends on a number of factors, including the variance of distributions and the visualization author's choice of axis scale as noted by previous work~\cite{Correll2020,Hofman2020,Witt2019}.

% \jessica{Cite Witt, Hofman, and Correll findings about the importance of axis limits to give a stronger rationale why we think distance matters.}

We investigate a scenario where distance heuristics lead to a predictable pattern of bias in order to measure how different visualization designs impact users' reliance on distance as a proxy for effect size.
% Imagine 
Users are shown charts depicting various effects on a fixed axis (Fig.~\ref{fig:distance_strategy}) such that when distributions have lower variance, visual distance between means is small regardless of effect size, but distances correspond to effect size more consistently at higher variance.
In this scenario, we expect that adding means to uncertainty visualizations leads users to underestimate effect size at lower variance. Conversely, adding means may reduce this underestimation bias at higher variance.
% While this prediction depends on fixing axis limits across charts, it gives us a behavioral signature to look for in order to determine whether users rely on a visual distance heuristic and discount uncertainty information. \jessica{omit prev sentence - intro is never a good place to cite caveats, and this sentence isn't necessary}

% Contribution statement
We contribute a pre-registered 
% mixed design 
experiment on Mechanical Turk 
% that uses such a scenario to investigate 
investigating how uncertainty visualization design impacts lay users' judgments and decisions from effect size.
We find that \textit{visualization designs which support magnitude estimation are not necessarily best suited as decision aids}. %\sout{and the best visualizations for the latter depend on chart characteristics.} 
Quantile dotplots lead to the least bias in magnitude estimation, 
% However, different visualization designs
but other visualizations lead to the least bias in decision-making.
On a fixed axis scale, densities without means support unbiased decisions at lower variance, and users show substantial bias with all visualizations at higher variance. 
Visualization effectiveness for decision-making depends on the level of variance in data relative to the axis scale.
% The best visualization for decision-making seems to depend on the span of distributions relative to axis length, which depends on both the level of variance in the data and axis scale. 
Adding means has a negligible impact on magnitude estimation, but in most cases it 
% biases users to make 
leads to
less utility-optimal decisions.

In a qualitative analysis of users' strategy descriptions, we find that \textit{few users apply the optimal strategy} for reading an uncertainty visualization when one exists. Instead, the majority of users appear to \textit{satisfice}~\cite{Simon1956} by using a small set of heuristics.
We find that the \textit{majority of users report relying on visual distance} between distributions regardless of uncertainty information, an observation that is consistent with the biases in our quantitative results.
We also find that many users switch between strategies. 
This suggests that many uncertainty visualizations may not be interpreted in ways that researchers and designers expect, and characterizing possible strategies may lead to design recommendations based on how users reason in practice.

%% file: 2_background.tex
\section{Background: Visualizing Uncertainty}
In communicating the results of statistical analysis, visualization authors commonly represent uncertainty as a range of possible values as recommended by numerous experts (e.g.,~\cite{Cumming2014,Manski2018b,taylor_guidelines_1994}). 
Other conventional uncertainty representations commonly used in statistical analysis include aggregate encodings of distributions such as boxplots~\cite{Tukey1977}, histograms~\cite{Pearson1895}, and densities~\cite{Spiegelhalter1999,Barrowman2003}.
Frequency-based uncertainty visualizations build on a large body of work suggesting that framing probabilities as frequencies of events improves statistical reasoning~\cite{Chance2000,Galesic2009,Gigerenzer1995,Hoffrage1998,Hogarth2011,Hullman2015,Kay2016,Kale2019-hops,Kim2019,Micallef2012,Fernandes2018}.
These include hypothetical outcome plots (HOPs)~\cite{Hullman2015}, which encode possible outcomes as frames in an animation, and quantile dotplots~\cite{Kay2016}, which quantize a distribution of possible outcomes and represent each quantile as a discrete dot.
A growing body of work suggests that lay and expert audiences commonly misinterpret interval representations of uncertainty~\cite{Belia2005,Soyer2012,Hoekstra2014} and that other uncertainty visualization formats such as gradient plots~\cite{Correll2014}, violin plots~\cite{Correll2014,Hullman2015}, HOPs~\cite{Hullman2015,Kale2019-hops}, and quantile dotplots~\cite{Kay2016,Fernandes2018} lead to more accurate interpretation and performance on various tasks. 

In our study, we compare two frequency-based visualizations, quantile dotplots and HOPs, with two more conventional uncertainty representations, intervals and densities.
By testing each with and without added means, we investigate the extent to which users of these uncertainty visualizations differ in their tendency to ignore uncertainty.

When chart users don't know how to interpret uncertainty, prior work~\cite{Hullman2015} suggests that they may substitute a judgment of the mean difference between distributions for more complicated judgments about the reliability of effects.
This \textit{visual distance heuristic} motivates design principles, for example, that the quantitative axis on a bar chart should always start at zero~\cite{Brinton1939,Huff1993}, or that axis scales should align visual distance with effect size~\cite{Witt2019}.
% or that visualization designers should choose axis scales which align visual distance with effect size~\cite{Witt2019}. 
Axis scale impacts the perceived importance of effect size regardless of chart type (e.g., lines versus bars) and despite attempts to signal that an axis does not start at zero (e.g., breaking the axis)~\cite{Correll2020}.
Rescaling the axis on a chart that displays inferential uncertainty (e.g., 95\% confidence intervals) to the scale implied by descriptive uncertainty (e.g., 95\% predictive intervals) can reduce bias in impressions of effect size~\cite{Hofman2020}.
% Rescaling the axis on a chart that displays inferential uncertainty (e.g., 95\% confidence intervals) to the scale implied by descriptive uncertainty (e.g., 95\% predictive intervals) can have a debiasing effect on impressions of effect size in standard scientific reports of results~\cite{Hofman2020}.
In our study, we investigate the visual distance heuristic by asking users to compare distributions with different levels of variance on a common scale (Fig.~\ref{fig:distance_strategy}).

% \jessica{Our background section is extremely short, that would be one section to fill out slightly, eg talk about the cloest prior work, eg original hops paper that suggested some evidence of a mean diff heuristic, witt paper maybe, hofman paper, fernandes paper on decision making from qdps, intervals, etc}

%% file: 3_method.tex
\section{Method}
We tested how adding means to different uncertainty visualizations impacts users’ estimates and incentivized decisions from effect size.

\subsection{Tasks \& Procedure}
Our task was like a fantasy sports game.
We showed participants charts comparing the predicted number of points scored by their team with and without adding a new player (e.g., Fig.~\ref{fig:distance_strategy}). 
Participants estimated the effect size of adding the new player and decided whether or not to pay to add the new player to their team. 

\textbf{Effect Size Estimation:}
We asked participants to estimate a measure of effect size called \textit{probability of superiority} or common language effect size~\cite{McGraw1992}: \textit{``How many times out of 100 do you estimate that your team would score more points with the new player than without the new player?''}
We elicited probabilities as ``times out of 100'' based on literature in statistical reasoning (e.g.,~\cite{Gigerenzer1995,Hoffrage1998}) suggesting that people reason more accurately with probabilities when they are framed as frequencies.
Probability of superiority, the percent of the time that outcomes for one group $A$ exceed outcomes for another group $B$, is a proxy for standardized mean difference $\frac{\mu_A - \mu_B}{\sigma_{A-B}}$~\cite{Coe2002,Cummings2011}, the difference between two group means relative to uncertainty in the estimates.
Using synthetic data (see Section 3.5), we evaluated bias in effect size estimates compared to a known ground truth.

\textbf{Intervention Decisions:}
We also asked participants to make binary decisions indicating whether they would \textit{``Pay for the new player,''} or \textit{``Keep [their] team without the new player.''}
On each trial, the participant's goal was to win an award worth \$3.17M, and they could pay \$1M to add a player to their team if they thought the new player improved their chances of winning enough to be worth the cost.
There were four possible payouts in each trial:
\begin{enumerate}[noitemsep,nolistsep]
    \item The participant won without paying for a new player (+\$3.17M).
    \item The participant paid for a new player and won (+\$2.17M).
    \item They failed to win without paying for a new player (\$0).
    \item The participant paid for a new player and failed to win (-\$1M). 
\end{enumerate}
The user could only lose money if they paid for the new player.\footnote{In pilot studies, we tested how framing outcomes as winning versus losing awards impacted user behavior and found that participants had greater preference for intervention when it was described as increasing the certainty of gains, consistent with prior work by Tversky and Kahneman~\cite{Kahneman1979,Tversky1981}.}
We set up the incentives for our task so that a risk-neutral chart user should pay for a new player only when effect size was larger than 74\% probability of superiority or Cohen’s d of 0.9, the average effect size in a recent survey of studies in experimental psychology~\cite{Schafer2019}.
This enabled us to evaluate intervention decisions compared to a utility-optimal standard.

\textbf{Feedback:}
At the end of each trial we told users whether or not their team scored enough points to win an award, using a Monte Carlo simulation to generate a win or loss based on the participant's decision. 
We split feedback into two tables. 
One showed the change in account value for the current trial. 
The other showed cumulative account value and how this translated into a bonus in real money. 
By showing probabilistic outcomes, instead of the expected value of decisions, feedback gave participants a noisy signal of how well they were doing, mirroring real-world learning conditions for decisions under uncertainty. 

\textbf{Payment:}
Participants received a \textit{guaranteed reward of \$1} plus a bonus of $\$0.08\cdot(account - \$150M)$, where \$0.08 per \$1M was the exchange rate from account value to real dollars, $account$ was the value of their fantasy sports account at the end of the experiment, and \$150M was a cutoff account value below which they receive no bonus. 
These values were carefully chosen to result in bonuses between \$0 and \$3, such that participants who guessed randomly and experienced unlucky probabilistic outcomes would receive no bonus, and participants who responded optimally would be guaranteed a bonus.

\textbf{User Strategies:}
To supplement our quantitative measures with qualitative descriptions of users' visual reasoning, at the end of each of the two block of trials, we asked users, \textit{``How did you use the charts to complete the task? Please do your best to describe what sorts of visual properties you looked for and how you used them.''}

\subsection{Formalizing a Class of Decision Problems}
Our decision task represents a class of decision problems where one makes a \textit{binary decision} about whether or not to invest in an intervention that changes the probability of an \textit{all-or-nothing outcome}. 
For example, this class of problems includes medical decisions about treatments that may save someone’s life or cure them of a disease, organizational decisions about hiring personnel to reach a contract deadline, and personal decisions such as paying for education to seek a promotion.
Previous decision-making literature examines similar problems in the context of salting the road in freezing weather~\cite{Joslyn2012,Joslyn2013}, voting in presidential elections~\cite{Westwood2019}, and willingness to pay for interventions in a fictional scenario~\cite{Hofman2020}.
The key similarity between these decision problems is that their incentive structures imply a \textit{common utility function}.

A utility function defines optimal (i.e., utility maximizing~\cite{vonNeumann1944}) decisions for a risk-neutral observer, providing a normative benchmark used to measure bias in decision-making. 
Comparing behavior to a risk-neutral benchmark is a common practice in judgment and decision-making studies~\cite{Baron2008}, often used to measure risk preferences~\cite{Weber1994} or attitudes that make a person more or less inclined to take action than they should be based on a cost-benefit analysis.
In the class of decision problems we investigate, the implied utility function depends on both the amount of money one stands to win or lose (e.g., the value of an award and the cost of a new player) and the effect size (e.g., the difference in team performance with versus without a new player).

Let $v$ be the value of an award.
Let $c$ be the cost of adding a new player to the team. 
The utility-optimal decision rule is to intervene if
\begingroup
\setlength\abovedisplayskip{3pt}
\setlength\belowdisplayskip{3pt}
    $$v \cdot \Pr(award|\neg player) < v \cdot \Pr(award|player) - c$$
\endgroup
where $\Pr(award|\neg player)$ is the probability of winning an award \textit{without} a new player, and $\Pr(award|player)$ is the probability of winning an award \textit{with} a new player.
Assuming a constant ratio between the value of the award and the cost of intervention $k = \frac{v}{c}$, we express the decision rule in terms of the difference between the probabilities of winning an award with versus without a new player: 
\begingroup
\setlength\abovedisplayskip{2pt}
\setlength\belowdisplayskip{2pt}
    $$\Pr(award|\neg player) + \frac{1}{k} < \Pr(award|player)$$
\endgroup

The threshold level of effect size above which one should intervene depends on the incentive ratio $k$ and the probability of a payout without intervention $\Pr(award|\neg player)$.
In our study, we fixed the incentives $k = 3.17$ and the probability of winning an award without a new player $\Pr(award|\neg player) = 0.5$ so that users would not have to keep track of changing incentives, and effect size alone was the signal that users should base decisions on.\footnote{In pilot studies, we tried manipulating $k$ and $\Pr(award|\neg player)$ and found that these changes had little impact on the effectiveness of different uncertainty visualizations for supporting utility-optimal decision-making. In light of prior work showing that Mechanical Turk workers do not respond to changes of incentives~\cite{Stoycheff2016}, we suspect that these manipulations might have an impact in real-world settings which is difficult to measure on crowdsourcing platforms.}
This enabled a controlled evaluation of how users translate visualized effect size into a sense of utility.
By modeling a functional relationship between effect size and utility, we go beyond prior work which either does not vary the effectiveness of interventions  (e.g.,~\cite{Joslyn2012,Joslyn2013,Westwood2019}) or examines only two levels of effect size as a robustness check for statistical tests (e.g.,~\cite{Hofman2020}).

\subsection{Experimental Design}
We assigned each user to one of four uncertainty visualization conditions at random, making comparisons of uncertainty visualizations between-subjects.
On each trial, users made a probability of superiority estimate and an intervention decision.
We asked users to make repeated judgments for two blocks of 16 trials each.
In one block, we showed the users visualizations with means added, and in the other block there were no means. 
We counterbalanced the order of these blocks across participants. 
Each of the 16 trials in a block showed a unique combination of ground truth effect size (8 levels) and variance of distributions (2 levels), making our manipulations of ground truth, variance, and adding means all within-subjects.
The order of trials in each block was randomized. 
In the middle of each block, we inserted an attention check trial, later used to filter participants who did not attend to the task. Users always saw an attention check at 50\% probability of superiority with means and at 99.9\% without means.
Hence, each participant completed 17 trials per block and 34 trials total. 
% \jessica{watch the wordiness! I imagine you could recover at least 10 lines, probably more from eliminating verbosity.}

\subsection{Uncertainty Visualization Conditions}
% \jessica{have you mentioned yet whether vis conditions are between participants or not? seems important context for reader as they read about conditions}
We evaluated visualizations intended to span a design space characterized by the visual salience of the mean, expressiveness of uncertainty representation, and discrete versus continuous encodings of probability. 
As described above, we showed four uncertainty visualization formats---intervals, hypothetical outcome plots (HOPs), density plots, and quantile dotplots---with and without separate (i.e., extrinsic) vertical lines encoding the mean of each distribution.
We expected that adding means would bias effect size estimates toward discounting uncertainty and that this effect would be most pronounced for uncertainty visualizations in which the mean is \textit{not} intrinsically salient. 

\textbf{Intervals:}
We showed users intervals representing a range containing 95\% of the possible outcomes (Fig.~\ref{fig:teaser}, left column). 
In the absence of a separate mark for the mean, the mean was not intrinsically encoded, and the user could only find the mean by estimating the midpoint of the interval.
% While the middle of the interval implicitly encoded the mean, this uncertainty encoding did not make the mean intrinsically salient. \jessica{you mean without mean added? Maybe say, We chose to test intervals because in the absence of a separate mark for the mean, the mean is not intrinsically encoded, and intervals are not very expressive .. ,}
Intervals were not very expressive of probability density since they only encoded lower and upper bounds on a distribution. 
% Numerous studies have shown that intervals are not the most effective uncertainty encoding~\cite{Correll2014,Hullman2015,Kale2019-hops} and that laypeople and scientists alike struggle to interpret them~\cite{Belia2005,Soyer2012,Hoekstra2014,Hofman2020}, yet they remain very common. \jessica{Don't we say this in related work}

\textbf{Hypothetical Outcome Plots (HOPs):}
We showed users animated sequences of strips representing 20 quantiles sampled from a distribution of possible outcomes (Fig.~\ref{fig:teaser}, left center column), matching the data shown in quantile dotplots.
Animations were rendered at \textit{2.5 frames per second with no animated transitions} (i.e., tweening or fading) between frames, looping every 8 seconds. 
We shuffled the two distributions of 20 quantiles using a 2-dimensional quasi-random Sobol sequence~\cite{Sobol1976} to minimize the apparent correlation between distributions.
Like intervals, HOPs did not make the mean intrinsically salient, as means were implicitly encoded as the average position of an ensemble of strips shown over time.
However, HOPs were more expressive of the underlying distribution than intervals and expressed uncertainty as frequencies of events, so they conveyed an experience-based sense of probability. 

\textbf{Densities:} 
We showed users continuous probability densities where the height of the area marking encoded the probabilities of corresponding possible outcomes on the \textit{x}-axis (Fig.~\ref{fig:teaser}, right center column).
Unlike intervals and HOPs, the mean was explicitly represented as the point of maximum mark height because distributions were symmetrical, so means were intrinsically salient.
Densities were also the most expressive of the underlying probability density function among the uncertainty visualizations we tested.
% \sout{, prior work found that users interpret probability density more accurately when it is discretized, i.e., quantile dotplots~\cite{Kay2016,Fernandes2018}. Thus, the expressiveness of densities may not translate into effective use. }

\textbf{Quantile Dotplots:}
We showed users dotplots where each of 20 dots represented a 5\% chance of a corresponding possible outcome on the \textit{x}-axis (Fig.~\ref{fig:teaser}, right column).
Like densities, because distributions were symmetrical and dots were stacked in bins to express this symmetry, the mean was explicitly represented as the point of maximum height and was thus intrinsically salient. 
% \jessica{My edits in here are aimed at 1) using more similar language between sections and 2) making clear the continuums we are talking about in intro to section}

\subsection{Generating Stimuli}
We generated synthetic data covering a range of effect size, so there were an equal number of trials where users should and should not intervene.
Recall that 50\% corresponded to a new player who did not improve the team’s performance at all, 100\% corresponded to a definite improvement in performance, and 74\% was the utility-optimal decision threshold.
We sampled eight distinct levels of ground truth probability of superiority, four values between 55\% and 74\% and four values between 74\% and 95\%, such that there are an equal number of trials above and below the utility-optimal decision threshold.
Prior work in perceptual psychology~\cite{Gonzalez1999,Zhang2012} suggests that the brain represents probability on a log odds scale.
For this reason, we converted probabilities into log odds units and sampled on this logit-transformed scale using linear interpolation between the endpoints of the two ranges described above. 
We added two attention checks at probabilities of superiority of 50\% and 99.9\%, where the decision task should have been very easy, to allow for excluding participants who were not paying attention.

To derive the visualized distributions from ground truth effect size, we made a set of assumptions.
We assumed \textit{equal and independent variances} for the distributions with and without a new player $\sigma_{team}^2$ such that $\sigma_{diff}^2 = 2\sigma_{team}^2$ 
where $\sigma_{diff}^2$ was the variance of the difference between distributions.
We tested two levels of variance, setting the standard deviation of the difference between distributions $\sigma_{diff}$ to a low value of 5 or a high value of 15. 
These levels produced distributions that looked relatively narrow or wide compared to the width of the chart, making visual distance between distributions an unreliable cue for effect size such that at low variance large effect sizes corresponded to distributions that looked close together.

We determined the distance between distributions, or mean difference $\mu_{diff}$, using the formula $\mu_{diff} = d\cdot\sigma_{diff}$
where $d$ were ground truth values as standardized mean differences (i.e., Cohen’s d~\cite{Coe2002,Cummings2011}).
The mean number of points scored without the new player was held constant $\mu_{without} = 100$, which corresponded to a 50\% chance of winning the award. 
We calculated mean for the team with a new player $\mu_{with} = \mu_{without} + \mu_{diff}$.
We rendered our chart stimuli using the parameters $\mu_{with}$, $\mu_{without}$, and $\sigma_{team}$ to define the two distributions on each chart.
% \jessica{Remind reader why we are doing all this}
Holding the chance of winning without a new player constant at 50\% (Fig.~\ref{fig:distance_strategy}, blue distributions) is an experimental control that enables us to compare a user's preference for new players across trials using a coin flip gamble as the alternative choice, which is common in judgment and decision-making studies~\cite{Baron2008}.  
% \jessica{these constraints simplify the user's task, might want to motivate why this is good}

\subsection{Modeling}
We wanted to measure how much users underestimate effect size in their probability of superiority responses, how much they deviate from a utility-optimal criterion in their decisions, and how sensitive they are to effect size for the purpose of decision-making.
To measure underestimation bias, we fit a linear in log odds model~\cite{Gonzalez1999,Zhang2012} to probability of superiority responses, and we derive \textit{slopes} describing users' responses as a function of the ground truth (Fig.~\ref{fig:llo_explainer}).
To measure bias and sensitivity to effect size in decision-making, we fit a logistic regression to intervention decisions, and we derive \textit{points of subjective equality} and \textit{just-noticeable differences} describing the location and scale of the logistic curve as functions of effect size (Fig.~\ref{fig:logistic_explainer}). 
% Following prior work~\cite{Gonzalez1999,Zhang2012}, both models assume that the user's internal representation of probabilities (i.e., probability of superiority and probability of winning) are on a log odds scale.

\subsubsection{Approach}
We used the brms package~\cite{Burkner2020} in R to build Bayesian hierarchical models for each response variable: probability of superiority estimates and decisions of whether or not to intervene. 
We started with simple models and gradually added predictors, checking the predictions of each model against the empirical distribution of the data.
This process of \textit{model expansion}~\cite{Gabry2019} enabled us to understand the more complex models in terms of how they differ from simpler ones. 

We started with a \textit{minimal model}, which had the minimum set of predictors required to answer our research questions, and built toward a \textit{maximal model}, which included all the variables we manipulated in our experiment. 
We specified the minimal and maximal models for each response variable in our preregistration.\footnote{
https://osf.io/9kpmb
% We include an anonymized preregistration in Supplemental Materials for review. There is a version with author names on OSF.
} 

Expanding models gradually helped us determine priors one-at-a-time. 
Each time we added a new kind of predictor to the model (e.g., a random intercept per participant), we honed in on weakly informative priors using prior predictive checks~\cite{Gabry2019}.
% (i.e., visualizing model predictions implied by sampling from the joint distribution of priors)
We centered the prior for each parameter on a value that reflected no bias in responses.
We scaled each prior to avoid predicting impossible responses and to impose enough regularization to avoid issues with convergence in model fitting. 
We documented priors and model expansion in Supplemental Materials.\footnote{
https://github.com/kalealex/effect-size-jdm
% [Github URL redacted for review]
}

\subsubsection{Linear in Log Odds Model}
% \jessica{there is extra white space in this section that could be reduced around equations}
We use the following model (Wilkinson-Pinheiro-Bates notation~\cite{Wilkinson1973,Burkner2020,Pinheiro2020}) for responses in the probability of superiority estimation task: 
% \matt{put group effects last in each submodel?}
\begingroup
\setlength\abovedisplayskip{3pt}
\setlength\belowdisplayskip{3pt}
    \begin{flalign*}
    \setlength\abovedisplayskip{0pt}
    \setlength\belowdisplayskip{0pt}
    \mathrm{logit}(response_{\Pr(S)}) \sim &\mathrm{Normal}(\mu, \sigma) \\[-2pt]
    \mu = &\mathrm{logit}(true_{\Pr(S)})*means*var*vis*order \\[-2pt]
        +& \mathrm{logit}(true_{\Pr(S)})*vis*trial \\[-2pt] 
        +& \big(\mathrm{logit}(true_{\Pr(S)})*trial + means*var\big|worker\big) \\[-2pt]
    \mathrm{log}(\sigma) =         &\mathrm{logit}(true_{\Pr(S)})*vis*trial \\[-2pt] 
        +& means*order \\[-2pt]
        +& \big(\mathrm{logit}(true_{\Pr(S)}) + trial\big|worker\big)
    \end{flalign*}
\endgroup
% \jessica{I would reduce space between lines slighly}
% \jessica{Pr(S) seems like slightly weird notation, like treating superiority as an event?} \alex{Probability of superiority as an event makes sense given how we elicited the judgment: ``How many times out of 100?''}
Where $response_{\Pr(S)}$ is the user's probability of superiority response, $true_{\Pr(S)}$ is the ground truth probability of superiority, $trial$ is an index of trial order, $means$ is an indicator for whether or not extrinsic means are present, $var$ is an indicator for low versus high variance, $vis$ is a dummy variable for uncertainty visualization condition, $order$ is an indicator for block order, and $worker$ is a unique identifier for each participant used to model random effects.
Note that there are submodels for the mean $\mu$ and standard deviation $\sigma$ of user responses.

\textbf{Motivation}:
We apply a logit-transformation to both $response_{\Pr(S)}$ and $true_{\Pr(S)}$, changing units from probabilities of superiority into log odds, because prior work suggests that the perception of probability should be modeled as linear in log odds (LLO)~\cite{Gonzalez1999,Zhang2012}.
% \jessica{shouldn't we motivate the log transform here?}
We model effects on both $\mu$ and $\sigma$ because we noticed in pilot studies that the spread of the empirical distribution of responses varies as a function of the ground truth, visualization design, and trial order.
However, we are most interested in effects on mean response.
The term $\mathrm{logit}(true_{\Pr(S)})*means*var*vis*order$ tells our model that the slope of the LLO model varies as a joint function of whether or not means were added, the level of variance, uncertainty visualization, and block order (i.e., all of these factors interacted with each other).
This enables us to answer our core research questions, while controlling for order effects.
The term $\mathrm{logit}(true_{\Pr(S)})*vis*trial$ models learning effects, so we isolate the impact of uncertainty visualizations.
In both submodels, we added within-subjects manipulations as random effects predictors as much as possible without compromising model convergence.
% The random effects terms in both submodels are the result of a process of model expansion where we systematically attempted to incorporate as many within-subjects manipulations as possible without compromising model convergence (see Supplemental Materials).

\subsubsection{Logistic Regression}
We use this model to make inferences about intervention decisions: 
% \matt{functions and distributions are typically set in upright type, e.g. $\mathrm{Bernoulli}(p)$ and $\mathrm{logit}(p)$. I would also consider a more direct naming of the outcome variable than $decision$. Perhaps $intervene$?}
\begingroup
\setlength\abovedisplayskip{3pt}
\setlength\belowdisplayskip{3pt}
    \begin{flalign*}
    intervene \sim &\mathrm{Bernoulli}(p) \\[-2pt]
    \mathrm{logit}(p) = &evidence*means*var*vis*order \\[-2pt]
        +& evidence*vis*trial \\[-2pt]
        +&\big(evidence*means*var + evidence*trial\big|worker\big)
    \end{flalign*}
\endgroup
Where $intervene$ is the user's choice of whether or not to intervene, $p$ is the probability that they intervene, and $evidence$ is a logit-transformation of the utility-optimal decision rule (see Section 3.2):
\begingroup
\setlength\abovedisplayskip{2pt}
\setlength\belowdisplayskip{2pt}
    $$evidence = \mathrm{logit}(\Pr(award|player)) - \mathrm{logit}(\Pr(award|\neg player) + \frac{1}{k})$$
\endgroup
This gives us a uniformly sampled scale of evidence where zero represents the utility-optimal decision threshold.
All other factors are the same as in the LLO model (see Section 3.6.2).

\textbf{Motivation}:
We logit-transform our evidence scale because internal representations of probabilities are thought to be on a log odds scale~\cite{Gonzalez1999,Zhang2012}, such that linear changes in log odds appear similar in magnitude.
% This evidence scale is perceptually uniform in theory~\cite{Gonzalez1999,Zhang2012,Varshney2013}. \jessica{unpack perceptually uniform in theory (especially the in theory part, what should the reader get from that statement?)} 
The term $evidence*means*var*vis*order$ tells our model that the location and scale of the logistic curve vary as a joint function of whether or not means were added, the level of variance, uncertainty visualization, and block order.
Mirroring an analogous term in the LLO model, this enables us to answer our core research questions, while controlling for order effects.
The term $evidence*vis*trial$ models learning effects.
As with the LLO model, we specify random effects per participant through model expansion by trying to incorporate as many within-subjects manipulations as possible.
% In both the LLO model and logistic regression, random effects are limited by identifiability, (i.e., whether we have enough trials per manipulation per participant for the model to estimate random effects parameters per participant).
% Adding random effects improves the fit of our models, however, we do not include them if doing so would require us to compromise on fixed effects which are more important for answering our research questions. 

\subsection{Derived Measures}
From our models, we derive estimates for three preregistered metrics that we use to compare visualization designs. 
% \jessica{reading the first part of this section, I felt like I was lacking an understanding of the measures we care about. Can we motivate these at least in a sentence or two earlier in section, at least in terms of what broad types of things we hope to learn, if not the specific measures? Without an idea of what measures we care about the model specs are hard to care about, it reads like a bunch of minute details without much context}

\begin{figure}[t]
    \centering
    \includegraphics[width=\columnwidth]{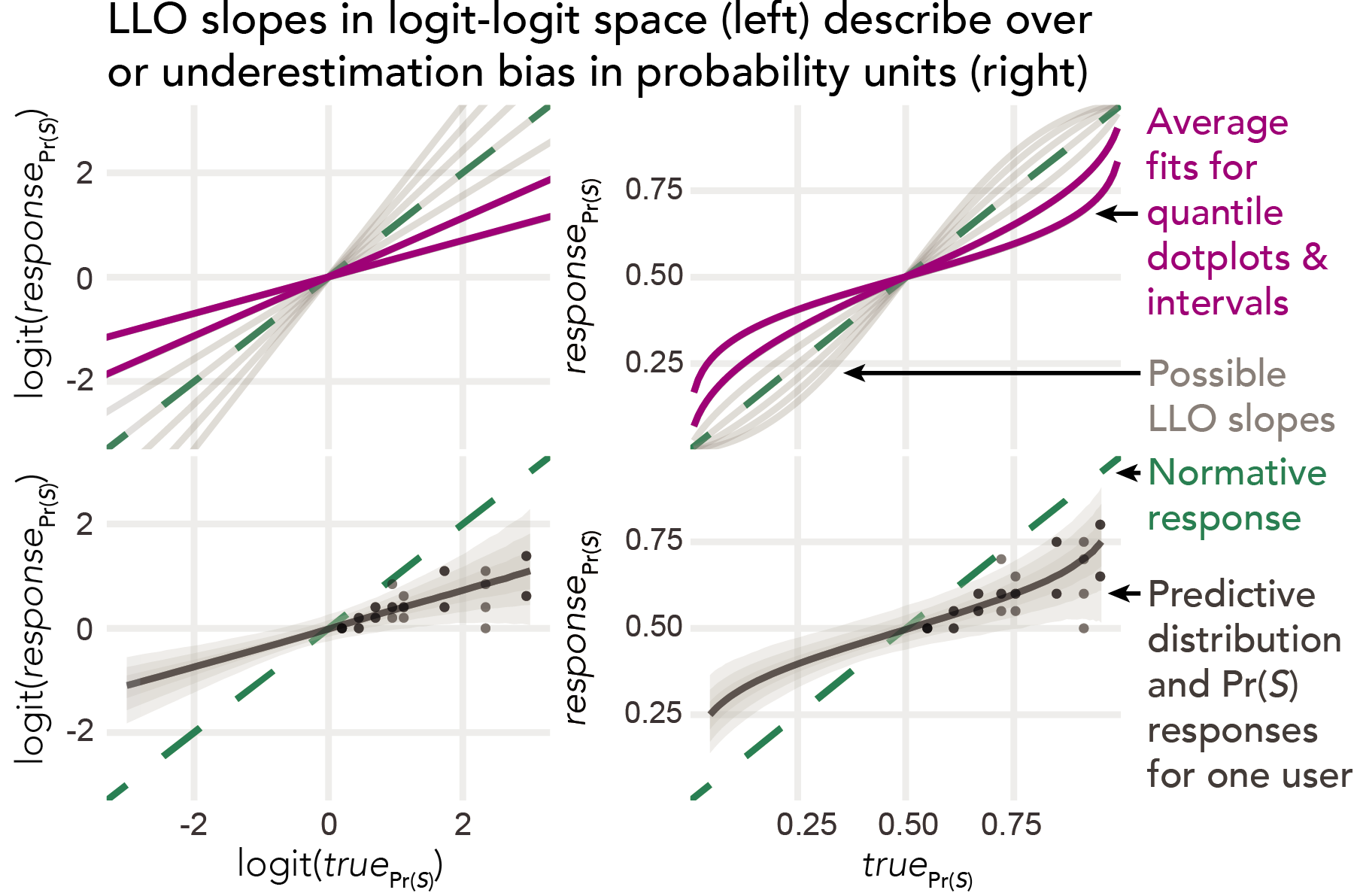}
    \setlength{\abovecaptionskip}{-10pt}
    \setlength{\belowcaptionskip}{-10pt}
    \caption{Linear in log odds (LLO) model: fits for average user of quantile dotplots and intervals compared to a range of possible slopes (top); predictive distribution and observed responses for one user (bottom).
    }
    \label{fig:llo_explainer}
\end{figure}

\begin{figure}[b]
    \vspace{-2mm}
    \centering
    \includegraphics[width=\columnwidth]{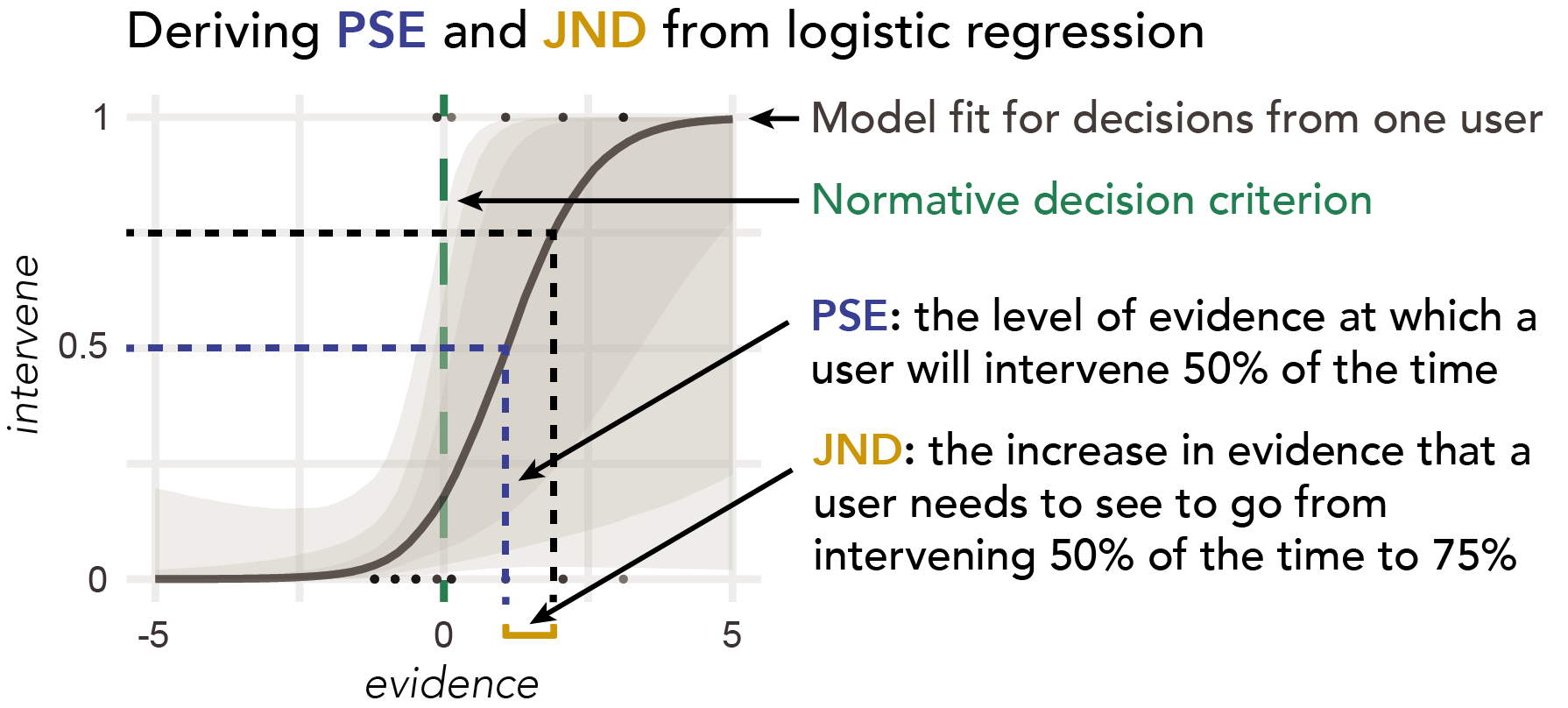}
    \setlength{\abovecaptionskip}{-10pt}
    \setlength{\belowcaptionskip}{-10pt}
    \caption{Logistic regression fit for one user. We derive point of subjective equality (PSE) and just-noticeable difference (JND) by working backwards from probabilities of intervention to levels of evidence.
    }
    \label{fig:logistic_explainer}
\end{figure}

\textbf{Linear in log odds (LLO) slopes} measure the degree of bias in probability of superiority $\Pr(S)$ estimation (Fig.~\ref{fig:llo_explainer}).
A slope of one indicates unbiased performance, and slopes less than one indicate the degree to which users underestimate effect size.\footnote{LLO slopes less than one represent bias toward the probability at the intercept, $\mathrm{logit^{-1}}(intercept)$, which is close to $\Pr(S) = 0.5$ in our study.}
We measure LLO slopes because they are very sensitive to the expected pattern of bias in responses, giving us greater statistical power than simpler measures like accuracy.
Specifically, LLO slope is the expected increase in a user's logit-transformed probability of superiority estimate, $\mathrm{logit}(response_{\Pr(S)})$, for one unit of increase in logit-transformed ground truth, $\mathrm{logit}(true_{\Pr(S)})$.
Using a linear metric (i.e., slope in logit-logit space) to describe an exponential response function in probability units comes from a theory that the brain represents probabilities on a log odds scale~\cite{Gonzalez1999,Zhang2012}.
The LLO model~\cite{Gonzalez1999,Zhang2012} can be thought of as a generalization of the cyclical power model~\cite{Hollands2000} that allows a varying intercept or a modification of Stevens' power law~\cite{Stevens1957} for proportions.
% We can think of the LLO model~\cite{Gonzalez1999,Zhang2012} as a modification of Steven's power law~\cite{Stevens1957} to handle proportions or as a special case  of the cyclical power model~\cite{Hollands2000} with reference probabilities of 0\% and 100\% \matt{I would more say that LLO can be thought of as a *generalization* (not special case) of the cyclical power model that allows the intercept to vary, where the cyclical power model is itself based on Stevens' power law (also note Stevens' not Steven's)}.

\textbf{Points of subjective equality (PSEs)} measure bias toward or against choosing to intervene in the decision task relative to a utility-optimal and risk-neutral decision rule (see Section 3.2).
PSEs describe the level of evidence at which a user is expected to intervene 50\% of the time (Fig.~\ref{fig:logistic_explainer}). 
A PSE of zero is utility-optimal, whereas a negative value indicates that a user intervenes when there is not enough evidence, and a positive value indicates that a user doesn't intervene until there is more than enough evidence. 
In our model, PSE is $\frac{-intercept}{slope}$ where $slope$ and $intercept$ come from the linear model in logistic regression.

\textbf{Just noticeable-differences (JNDs)} 
measure sensitivity to effect size information for the purpose of decision-making (Fig.~\ref{fig:logistic_explainer}). 
They describe how much additional evidence for the effectiveness of an intervention a user needs to see in order to increase their rate of intervening from 50\% to about 75\%.
A JND in evidence units is a difference in the log probability of winning the award with the new player.
We chose this scale for statistical inference because units of log stimulus intensity are thought to be approximately perceptually uniform~\cite{Stevens1957,Varshney2013}.
In our model, JND is $\frac{\mathrm{logit}(0.75)}{slope}$ where $slope$ is the same as for PSE.

\subsection{Participants}
We recruited users through Amazon Mechanical Turk.
Workers were located in the US and had a HIT acceptance rate of 97\% or more.
Based on the reliability of inferences from pilot data, we aimed to recruit 640 participants, 160 per uncertainty visualization.
We calculated this target sample size by assuming that variance in posterior parameter estimates would shrink by a factor of roughly $\frac{1}{\sqrt{n}}$ if we collected a larger data set using the same interface.
Since we based our target sample size on between-subjects effects (e.g., uncertainty visualization), our estimates of within-subjects effects (e.g., adding means) were very precise.

We recruited 879 participants.
After our preregistered exclusion criterion that users needed to pass both attention checks, we slightly exceeded our target sample size with 643 total participants.
However, we had issues fitting our model for an additional 21 participants, 17 of whom responded with only one or two levels of probability of superiority and 4 of whom had missing data.
After these non-preregistered exclusions, our final sample size was 622 (with block order counterbalanced). 
All participants were paid regardless of exclusions, on average receiving \$2.24 and taking 16 minutes to complete the experiment.

\subsection{Qualitative Analysis of Strategies}
Using the two strategy responses we elicited from each user, we conducted a qualitative analysis to characterize users' visual reasoning strategies based on heuristics they used with different visualization designs (with and without means) and whether they switched strategies.

The first author developed a bottom-up open coding scheme for how users described their reasoning with the charts.
Since some responses were uninformative about what visual properties of the chart a user considered (e.g., \textit{``I used the charts to estimate the value added by the new player.''}), we omitted participants for whom both responses were uninformative from further analysis.
Excluding 180 such participants resulted in a final sample of 442 for our qualitative analysis.

We used our open codes to develop a classification scheme for strategies based on what visual features of charts users mentioned, whether they switched strategies, and whether they were confused by the chart or task.
We coded for the following uses of visual features:
\begin{itemize}[noitemsep,nolistsep]
    \item \textbf{Relative position} of distributions
    \item \textbf{Means}, whether users relied on or ignored them
    \item \textbf{Spread} of distributions, whether users relied on variance, ignored it, or erroneously preferred high or low variance
    \item \textbf{Reference lines}, whether users relied on imagined or real vertical lines (e.g., the annotated decision threshold in Fig.~\ref{fig:teaser} \&~\ref{fig:distance_strategy})
    \item \textbf{Area}, whether users relied on the spatial extent of geometries
    \item \textbf{Frequencies}, whether users of quantile dotplots or HOPs relied on frequencies of dots or animated draws
\end{itemize}
Thus, we generated a spreadsheet of quotes, open codes, and categorical distinctions which enabled us to provide aggregate descriptions of patterns and heterogeneity in user strategies.
% }

% \begin{itemize}
%     \vspace*{-5pt}\item \textbf{Relative position} of distributions
%     \vspace*{-5pt}\item \textbf{Means}, whether users relied on or ignored them
%     \vspace*{-5pt}\item \textbf{Spread} of distributions, whether users relied on variance, ignored it, or erroneously preferred high or low variance
%     \vspace*{-5pt}\item \textbf{Reference lines}, whether users relied on imagined or real vertical lines (e.g., the annotated decision threshold)
%     \vspace*{-5pt}\item \textbf{Area}, whether users relied on the spatial extent of geometries
%     \vspace*{-5pt}\item \textbf{Frequencies}, whether users of quantile dotplots or HOPs relied on frequencies of dots or animated draws
% \end{itemize}
% \vspace*{-5pt} Thus, we generated a spreadsheet of quotes, open codes, and categorical distinctions which enabled us to provide aggregate descriptions of patterns and heterogeneity in user strategies.

%% file: 4_results.tex
% full page figure to start results
\includepdf{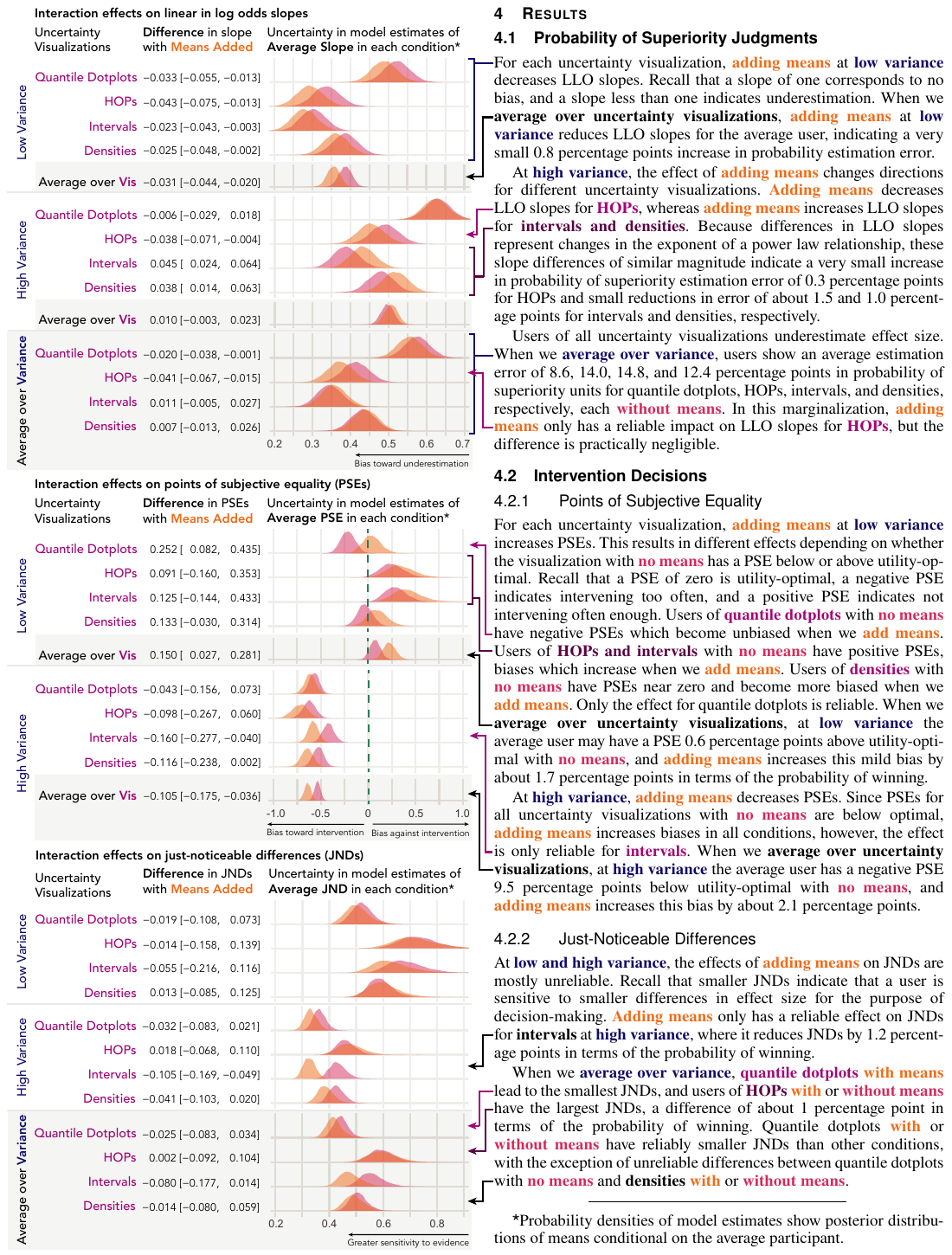}
% \jessica{results figure looks good but some of the arrows get confusing, eg sometimes there are arrowheads, sometimes not, in 4.2.1 and 4.2.2 a few are ambiguous} 

% be sure to include cites (if any)
% \nocite{}

% correct section counters
\setcounter{section}{4}
\setcounter{subsection}{2}

\subsection{Discussion}
Among the uncertainty visualizations we tested, quantile dotplots lead to the least biased probability of superiority estimates.
This is not surprising given previous work (e.g.,~\cite{Gigerenzer1995,Hoffrage1998,Hullman2015,Kale2019-hops,Kay2016}) showing that frequency-based visualizations are effective at conveying probabilities. 
However, it is surprising that users do not perform reliably differently with frequency-based HOPs than with intervals or densities.
% , each with or without means added. 
HOPs directly encode probability of superiority by how often the draws from the two distributions change order, whereas in all other conditions users would need to calculate effect size analytically from visualized means and variances to arrive at the ``correct'' inference, although we doubt that users engage in such explicit mathematical reasoning. 
In Section 5, we present descriptive evidence of heuristics that users employ with different visualization designs, which helps to explain these results.

In most cases, the small effects on LLO slopes when adding means to uncertainty visualizations are probably negligible.
However, they are consistent with the pattern of behavior we expect if users rely on visual distance between distributions as a proxy for effect size. 
When variance is lower relative the axis scale, distances between distributions look small even for large effects (Fig.~\ref{fig:distance_strategy}, top), and users tend to underestimate effect size \textit{more} when means are added. 
When variance is higher relative the axis scale, distances between distributions roughly correspond to effect size (Fig.~\ref{fig:distance_strategy}, bottom), and users tend to underestimate effect size \textit{less} when means are added, at least for densities and intervals. 
% \matt{it took me a few tries to parse this sentence}

Our results suggest that the best visualization design for utility-optimal decision-making probably depends on the level of variance relative to the axis scale.
% For promoting utility-optimal decisions, our results suggest that the best visualization design probably depends on the level of variance as it interacts with \matt{maybe say "relative to" instead of "as it interacts with"? Had to re-parse this sentence because I first read "as" as "because"} the axis scaling.
At lower variance, when multiple levels of variance are shown on a common scale, densities without means or quantile dotplots with means lead to the least bias in decisions.
At higher variance, users are biased toward intervening in all conditions, and both densities without means and intervals without means lead to the least bias.
The impact of means also depends on variance and axis scaling, such that when we average across uncertainty visualizations, adding means exacerbates biases that exist when means are absent.
The effect of variance on PSEs (see Supplemental Materials) is large, such that users intervene more often at higher variance than at lower variance.
One possible explanation for this is that users rely on distance between distributions as a proxy for effect size and make decisions as if effects are larger when distributions are further apart (Fig.~\ref{fig:distance_strategy}).
% \alex{We repeat this future work statement in the General discussion. Maybe omit.}
% Future work should investigate bias in decision-making over a gradient of variances visualized on a common scale, including charts with heterogeneous variances, as this would enable more exhaustive design recommendations about how visualizations for decision aids should be designed as a function of the span of distributions relative to the axis. 
% \matt{I like this para! good job acknowledging the complexity of it}
% \jessica{what about not a common scale? are we not recommending bc other papers do that? added the heterogenous phrase which might take care of my question}

% \alex{dial back the overclaiming here. mention the effect of variance and the possibility of a gradation of bias in between levels}
% While the impact of adding means on decision quality is small, our results suggest that means are unhelpful for promoting utility-optimal decisions. The one exception is that for quantile dotplots at low variance, adding means reduces bias in decision-making, however user performance is no less biased when using densities without means at low variance. At high variance, intervals with no means may be less biased than densities with no means, but the difference in performance is not reliable. Together this results suggest that densities without means promote decisions with the least bias in realistic use cases where distributions with heterogeneous variances are displayed on a common axis (e.g., in faceted small multiples).

Reported effects of visualization design on JNDs may not be practically important. 
All differences in JNDs between visualization designs are smaller than the difference between high versus low variance (see Supplemental Material). 
Smaller JNDs at high variance may reflect the fact that our high variance charts use white space more efficiently.

\subsection{Comparing Magnitude Estimation \& Decision-Making}
\begin{wrapfigure}{r}{0.34\columnwidth}
    \vspace*{-5mm}
    \centering
    \includegraphics[width=0.33\columnwidth]{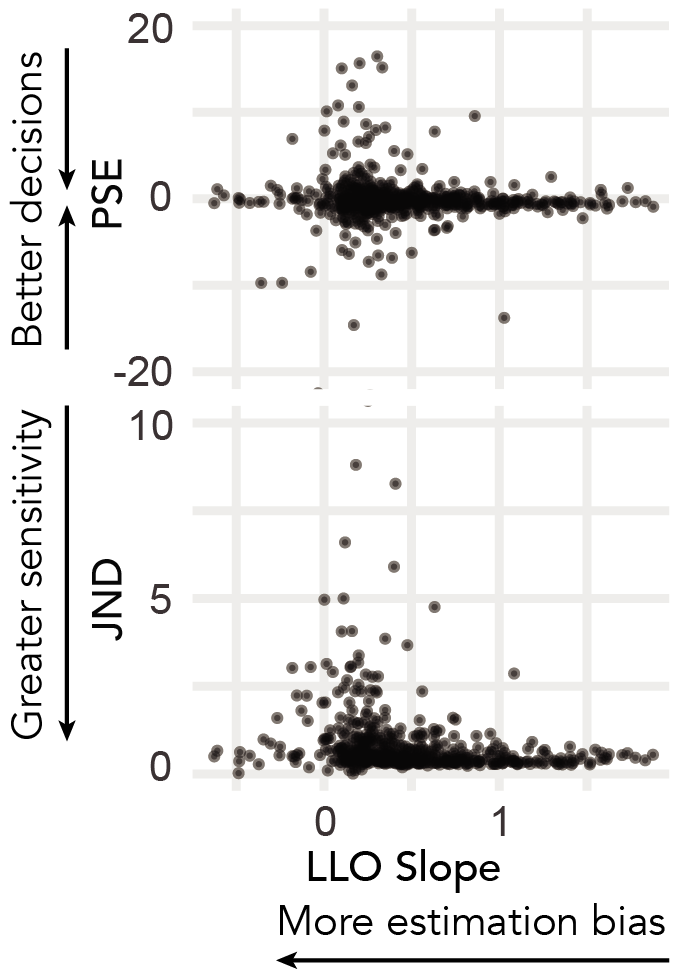}
    \setlength{\abovecaptionskip}{4pt}
    \caption{PSEs and JNDs vs LLO slopes per user.}
    \label{fig:decoupling}
    \vspace*{2mm}
    \centering
    \includegraphics[width=0.33\columnwidth]{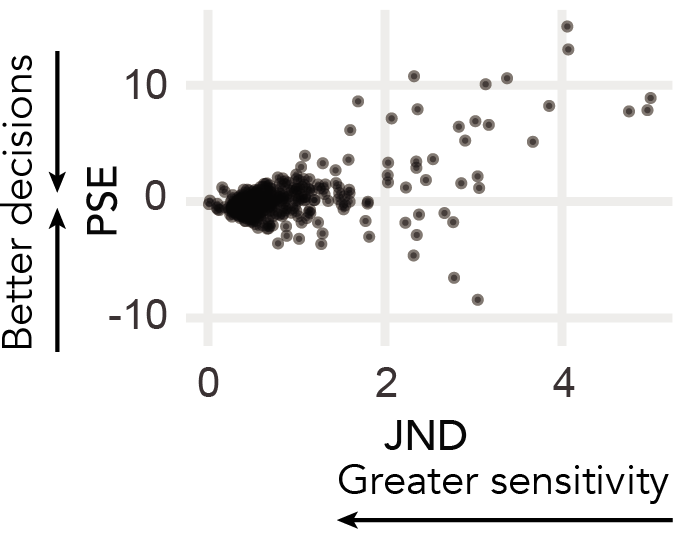}
    \setlength{\abovecaptionskip}{-7pt}
    \setlength{\belowcaptionskip}{-10pt}
    \caption{JNDs vs PSEs.}
    \label{fig:jnd-pse}
\end{wrapfigure}
Different visualization designs lead to the best performance on our magnitude estimation and decision-making tasks.
To explore this decoupling of performance across tasks, we calculate average posterior estimates of our derived measures---LLO slope, PSE, and JND---for each individual user and compare them.
Figure~\ref{fig:decoupling} shows that many individuals who are poor at magnitude estimation (i.e., LLO slopes below one) do well on the decision task (i.e., PSEs and JNDs near zero). 

One possible explanation for this decoupling of performance on our two tasks is that users may rely on different heuristics to judge the same data for different purposes.
This is consistent with Kahneman and Tversky's~\cite{Kahneman1979} distinction between \textit{perceiving} the probability of an event to be $p$ and \textit{weighting} the probability of an event in decision-making as $\mathrm{\pi}(p)$, which suggests that decision weights reflect preferences based on probabilities and risk attitudes~\cite{Weber1994}.
Recent work in behavioral economics~\cite{Khaw2017} suggests that biases in decision-making are partially attributable to imprecision in an individual's subjective perception of numbers (i.e., ``number sense'').
Since JNDs reflect the precision of perceived effect size implied by one's decisions and PSEs represent bias in decision-making, we can investigate this relationship within individual users in our study (Fig~\ref{fig:jnd-pse}).
In agreement with prior work, we see that greater sensitivity to effect size for decision-making (i.e., JNDs close to zero) predicts more utility-optimal decisions (i.e., PSEs close to zero). 
Although, based on the decoupling of LLO slopes and JNDs, it also seems clear that a user's internal sense of effect size is not necessarily identical when they use the same information for different tasks.
We should be mindful that perceptual accuracy may not feed forward directly into decision-making.

%% file: 5_strategies.tex
\section{Visual Reasoning Strategies}
We use qualitative analysis of reported strategies to identify ways that users judge effect size by comparing distributions, giving us a vocabulary for how visualization design choices impact their interpretations. 
% \jessica{kind of want to move this sentence up out of subsection bc I've always been taught its bad to start any level section without some text. Also, given that we have a little extra space, a figure here might be nice}

\subsection{Prevalent Strategies}
The strategies we identify are not mutually exclusive. We count a user as employing a strategy if they mention it in either of their responses.
% \jessica{some missing info here about how you classify users as doing something - in either of their strategy responses? What if its in one only, do you still say things like 'most users'?}

\textbf{Only Distance:}
About 62\% of users (275 of 442) rely on ``how far to the right'' the red distribution is compared to the blue one \textit{without mentioning that they incorporate the variance of distributions into their judgments} (Fig.~\ref{fig:distance_strategy}).
Roughly 69\% of these users (190 of 275) describe making a gist estimate of distance between distributions, with 46\% (126 of 275) saying they rely on the mean difference specifically, and 13\% (36 of 275) saying they rely on both gist distance and mean difference.
Strategies which involve only the distance between distributions should result in a large bias toward underestimating effect size, which is what we see in our aggregated quantitative results. 
% \jessica{do we see this in general}

\textbf{Distance Relative to Variance:}
Only about 8\% of users (35 of 442) mention that their interpretations of distance depend on the spread of distributions, suggesting that perhaps very few untrained users are sensitive to the impact of variance on effect size.
% Only about 8\% of users (35 of 442) seem to understand that interpretations of distance between distributions should depend on the spread of the distributions, suggesting that few are sensitive to the impact of variance on effect size. \matt{it strikes me there's a difference between people "seeming to understand it" versus "mention it"? Like, perhaps there are people who understand it but didn't happen to mention it in their response? I could see some people in the previous category being in that bucket --- this sentence feels very "absense of evidence taken as evidence of absence" and could be softened a bit}
If users estimate standard deviation and mean difference between distributions, they could use this information to calculate effect size analytically.
However, we think it is far more likely that these users judge the distance between distributions relative to the spatial extent of uncertainty visualizations, which should result in underestimation bias which is similar to but less pronounced than with judgments of \textit{only distance}.

\begin{wrapfigure}{r}{0.34\columnwidth}
    \vspace*{-4mm}
    \centering
    \includegraphics[width=0.33\columnwidth]{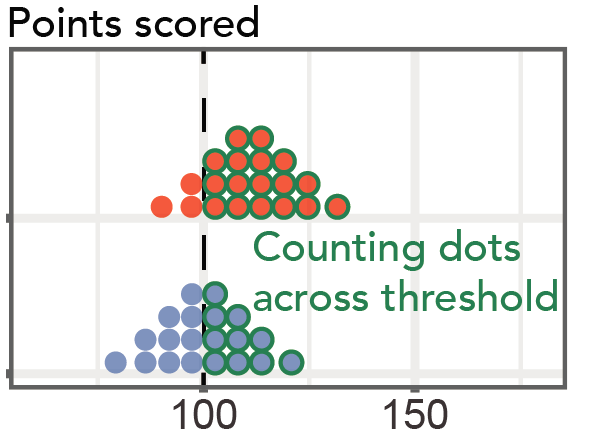}
    \setlength{\abovecaptionskip}{-5pt}
    \setlength{\belowcaptionskip}{-7pt}
    \caption{Cumulative probability strategy with quantile dotplots.}
    \label{fig:cumulative_p_strategy}
    \vspace*{3mm}
    \centering
    \includegraphics[width=0.33\columnwidth]{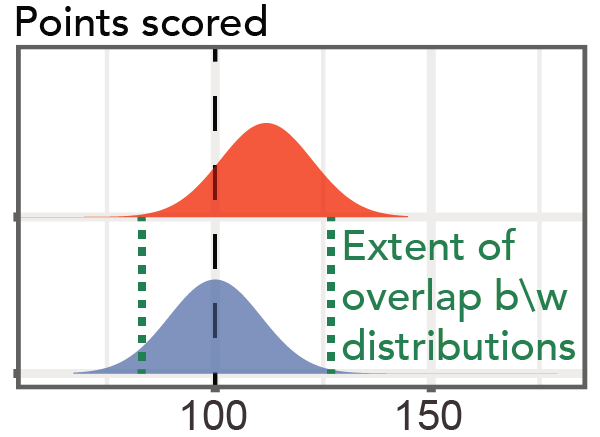}
    \setlength{\abovecaptionskip}{-5pt}
    \setlength{\belowcaptionskip}{-7pt}
    \caption{Overlap strategy with densities.}
    \label{fig:overlap_strategy}
\end{wrapfigure}

\textbf{Cumulative Probability:}
A substantial 36\% of users (160 of 442) estimate the cumulative probability of winning the award with and/or without the new player.
This strategy involves judging the distance, proportion of area, or frequency of markings across the threshold number of points to win (e.g., Fig.~\ref{fig:cumulative_p_strategy}).
These users may be confusing cumulative probability of winning the award, which is the best cue in the decision task, with probability of superiority (i.e., probability that team does better with the new player than without), which is what we ask for in the estimation task. 
However, since the probability of winning increases monotonically with probability of superiority, this strategy should theoretically result in milder underestimation bias than distance-based strategies. 

% \begin{wrapfigure}{r}{0.34\columnwidth}
%     \vspace*{-1mm}
%     \centering
%     \includegraphics[width=0.33\columnwidth]{figures/overlap_strategy-01.png}
%     \setlength{\abovecaptionskip}{-5pt}
%     \setlength{\belowcaptionskip}{-7pt}
%     \caption{Overlap strategy with densities.}
%     \label{fig:overlap_strategy}
% \end{wrapfigure}

\textbf{Distribution Overlap:}
About 7\% of users (31 of 442) describe judging the overlap between distributions.
While similar to distance-based strategies, users conceptualize this strategy in terms of area rather than the gap between distributions (Fig.~\ref{fig:overlap_strategy}).
For example, one user said they use HOPs \textit{``only to see how much of an overlap [there is] between the two areas,''} suggesting that they imagine contours of distributions over the sets of animated draws.
This strategy probably results in underestimation bias similar to judging \textit{distance relative to variance}.

\begin{figure}[h]
    \vspace*{-1mm}
    \centering
    \includegraphics[width=\columnwidth]{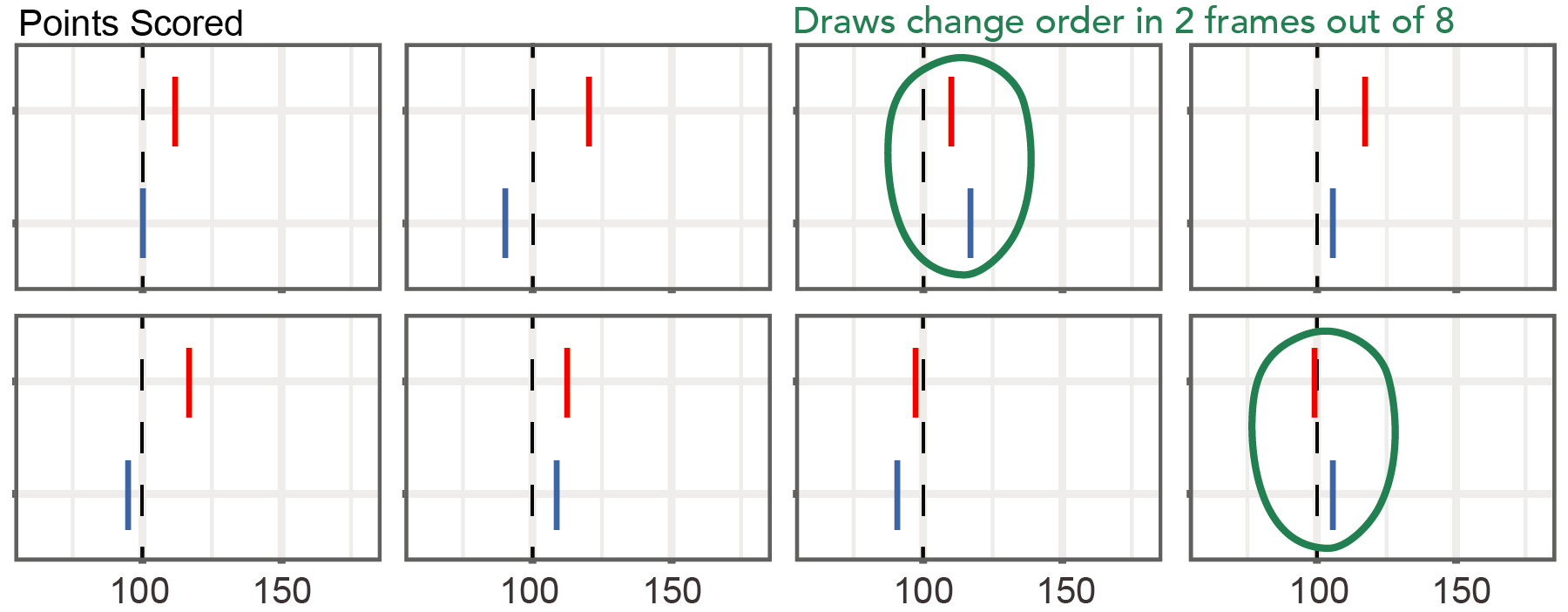}
    \setlength{\abovecaptionskip}{-10pt}
    \setlength{\belowcaptionskip}{-5pt}
    \caption{Frequency of draws changing order strategy with HOPs.}
    \label{fig:draws_changing_order_strategy}
\end{figure}

\textbf{Frequency of Draws Changing Order:}
This strategy is only relevant to the HOPs condition, where only about 16\% of users (19 of 121) employed it.
It involves judging the number of animated frames in which the draws from the two distributions switch order (Fig.~\ref{fig:draws_changing_order_strategy}).
This is the best way to estimate probability of superiority from HOPs~\cite{Hullman2015}.
If we think of the user as accumulating information across frames, the precision of their inference is mostly limited by the number of frames they watch.
For example, in Figure ~\ref{fig:draws_changing_order_strategy} red scores higher than blue in 6 of the 8 frames, and watching only 8 frames limits the precision of this inference to increments of $\frac{1}{8}$.
The fact that only a handful of HOPs users employ this strategy helps to explain why the performance of HOPs users is worse than expected.

\textbf{Switching Strategies:}
A substantial 29\% of users (129 of 442) switch between strategies in the middle of the task.
For example, one user of intervals without means described a mix of \textit{cumulative probability} and \textit{distribution overlap} strategies: \textit{``If the red [distribution] was completely past the dotted line then I would buy the new player no matter what. If there were overlaps with blue I would just risk assess to see if it was worth it to me or not.''} 
While more of a meta-strategy, our observation that a significant proportion of users switch is important because it suggests that judgment processes involved in graphical perception may not be consistent within each user.

% table of strategy frequecies per uncertainty vis
\begin{table}[b]
\vspace*{-3mm}
\setlength\tabcolsep{3pt} % default value: 6pt
\captionsetup{justification=centering,skip=0.25\baselineskip}
\begin{threeparttable}
    \caption{Frequency of strategies used per uncertainty visualization.}
    \label{tab:straties}
    \begin{tabularx}{\columnwidth}{@{} L RRRR|R @{}}
    \toprule 
    Strategy & Intervals & HOPs & Densities & Dotplots & Overall \\
    \midrule
    Distance & 73 & 77 & 61 & 64 & 275 \\
    Rel. to Var. & 11 & 9 & 10 & 5 & 35 \\
    Cumulative & 34 & 50 & 30 & 46 & 160 \\
    Overlap & 17 & 2 & 9 & 3 & 31 \\
    Draw Order & 0 & 19 & 0 & 0 & 19 \\
    Switching & 35 & 48 & 23 & 23 & 129 \\
    \hline
    Total & 112 & 121 & 99 & 110 & 442 \\
    \bottomrule
    \end{tabularx}
\end{threeparttable}
\end{table}
% \begin{table}[h]
%     \centering
%     \begin{tabular}{ |p{0.18\columnwidth}|p{0.11\columnwidth}|p{0.08\columnwidth}|p{0.12\columnwidth}|p{0.11\columnwidth}|p{0.09\columnwidth}| } 
%      \hline
%      Strategy & Intervals & HOPs & Densities & Dotplots & Overall \\
%      \hline
%      Distance & 73 & 77 & 61 & 64 & 275 \\
%      \hline
%      Rel. to Var. & 11 & 9 & 10 & 5 & 35 \\
%      \hline
%      Cumulative & 34 & 50 & 30 & 46 & 160 \\
%      \hline
%      Overlap & 17 & 2 & 9 & 3 & 31 \\
%      \hline
%      Draw Order & 0 & 19 & 0 & 0 & 19 \\
%      \hline
%      Switching & 35 & 48 & 23 & 23 & 129 \\
%      \hline
%      Total & 112 & 121 & 99 & 110 & 442 \\
%      \hline
%     \end{tabular}
% \end{table}

\subsection{Impacts of Visualization Design Choices}
Users rely on visual features (Section 3.9) and strategies (Section 5.1) to varying degrees depending on visualization design (Table~\ref{tab:straties}). 
% \jessica{I am kind of disliking the use of 'we' equating us with visualization designers. Most of these sentences can be rephrased to be less colloquial, eg to varying degrees based on the visualization design}
% Recall that strategies are not mutually exclusive and that users often switch between them.

\textbf{Intervals:}
Roughly 75\% of intervals users (85 of 112) rely on \textit{relative position} as a visual cue for effect size compared to 69\% with densities (68 of 99), 61\% with HOPs (74 of 121), and 59\% with quantile dotplots (65 of 110). 
%  \jessica{trying to make this easier to read, should try to do more of this in this section so the reader doesn't have to keep scanning between text and ()}
Of intervals users who look at \textit{relative position}, about 87\% (74 of 85) employ an \textit{only distance} strategy, while only about 13\% (11 of 85) judge \textit{distance relative to variance} .
In other words, only about 10\% of intervals users (11 of 112) incorporate variance into their judgments of distance.
About 28\% of intervals users (31 of 112) report looking at \textit{area}, with about 55\% of these users (17 of 31) employing a \textit{distribution overlap} strategy.

\textbf{HOPs:}
About 61\% of HOPs users (74 of 121) look at \textit{relative position} to judge effect size.
Of HOPs users who rely on \textit{relative position}, merely 3\% (2 of 74) use a \textit{distance relative to variance} strategy.  
However, looking at \textit{relative position} is not mutually exclusive with looking at \textit{frequency} of draws, which 45\% of HOPs users (54 of 121) rely on as a visual feature.
Among HOPs users who rely on frequencies, about 69\% (37 of 54) employ a \textit{cumulative probability} strategy, while about 35\% (19 of 54) rely on the optimal strategy of counting the \textit{frequency of draws changing order}.
Roughly 40\% of HOPs users (48 of 121) mention \textit{switching strategies} compared to 31\% with intervals (35 of 112), 23\% with densities (23 of 99), and 21\% with quantile dotplots (23 of 110).
Among HOPs users who switch strategies, about 81\% (39 of 48) rely on the mean as a cue.
Strategy switching involves the mean for about 30\% of HOPs users who rely on \textit{relative position} (22 of 74) compared to 43\% of HOPs users who rely on \textit{frequency} (23 of 54).
That most HOPs users rely on \textit{relative position}, and that those who do rely on \textit{frequency} are more likely to switch to or from relying on the mean, helps to explain poor performance with HOPs.
% why performance with HOPs is lower than expected.

\textbf{Densities:}
About 69\% of densities users (68 of 99) rely on \textit{relative position} as a visual cue.
Of densities users who look at \textit{relative position}, only about 13\% (9 of 68) employ a \textit{distance relative to variance} strategy.
As one might expect, a substantial 36\% of densities users (36 of 99) rely on \textit{area} as a cue, compared to 10\% of quantile dotplots users (11 of 110).
Among densities users who rely on \textit{area}, about 53\% (19 of 36) employ a \textit{cumulative probability} strategy, while about 28\% (10 of 36) employ a \textit{distribution overlap} strategy. 
Interestingly, about 27\% of densities users (27 of 99) mention relying on the \textit{spread} of distributions as a cue, more than the 21\% of users with intervals (24 of 112), 21\% with HOPs (25 of 121), and 10\% with quantile dotplots (11 of 110) who report relying on the same cue.

\textbf{Quantile Dotplots:}
Roughly 59\% of quantile dotplots users (65 of 110) describe looking at \textit{relative position} to judge effect size, similar to 61\% of users with HOPs (74 of 121) and less than the 69\% of densities users (68 of 99) and 76\% of intervals users (85 of 112) who report using the same cue.
Merely 6\% of quantile dotplots users who rely on \textit{relative position} (4 of 65) employ a \textit{distance relative to variance} strategy.
37\% of quantile dotplots users (41 of 110) rely on \textit{frequency} as a visual cue by counting dots.
About 81\% of quantile dotplots users who rely on \textit{frequency} (33 of 41) employ a \textit{cumulative probability} strategy. 

\textbf{Adding Means:}
A substantial 35\% of users (155 of 442) describe relying on the mean as a cue for effect size. 
If we split users based on whether or not they start the task with means, about 31\% of users (67 of 218) \textit{switch strategies} when means are added to the charts halfway through the task, compared to 10\% (23 of 224) who switch strategies when means are removed. 
This asymmetry in strategy switching suggests that means are ``sticky'' as a cue: 
% If a user starts relying on the mean they are less likely to switch to a different strategy than they are to switch from a different strategy to relying on means.
% \jessica{explain what you mean in previous sentence in more detail}
Among the 15\% of users (67 of 442) who start with and rely on means, about 66\% (44 of 67) attempt to visually estimate means \textit{after means are removed from charts}, almost twice as many as the 34\% (23 of 67) who switch to relying on other cues. 
However, the impact of adding means on performance depends on what other strategies a user is switching between.
Among the 20\% of users (90 of 442) who rely on means and switch strategies, about 44\% (40 of 90) just incorporate the mean into judgments of \textit{relative position}  without relying on other visual cues. 
Other groups of users switch between relying on means and less similar visual cues, with 34\% (31 of 90) also mentioning \textit{frequency} and 12\% (11 of 90) mentioning \textit{area}.
That many users switch between relying on \textit{relative position} and means, and that strategies are heterogeneous, helps to explain why the average impact of means on performance is small in our results. 

%% file: 6_discussion.tex
\section{General Discussion}
% \alex{Frame the summary points in this section as high-level take-aways for designers (and relevant caveats): 
% 1. Use quantile dotplots for perceptually accurate distributional comparisons (dual tasks may have led to relative underestimate of performance with HOPs); 
% 2. Densities and without means seem to support the best decision-making (effectiveness seems to depend on variance and axis scale in complex ways that need to be addressed by future work);
% 3. Adding means biases users toward underestimating effect size and making less utility optimal decisions (In most cases these effects are negligible, athough we may be underestimating them because... The biasing effect of adding means on decision-making seems to depend on variance and axis scale, such that this effect may disappear for specific combinations of variance and axis scale.)
% 4. Users will rely on distance between distributions as a proxy for effect size, so note when this will be misleading and try to encourage more optimal strategies (We may overestimate the extent to which chart in general users will rely on a distance heuristic in other situations since they also rely on other strategies, and specific design choices, e.g., adding reference lines, may impact users' tendency to rely on these strategies. )
% }

Our results suggest that \textit{design guidelines for visualizing effect size} should depend on the user's task, the variance of distributions, and design choices about axis scales.
To provide concrete design guidelines while acknowledging the inherent complexity of our results, we present high-level take-aways for designers alongside relevant caveats.

\textbf{Quantile dotplots support the most perceptually accurate distributional comparisons,} at least among the visualization designs we tested.
\textbf{Caveat:} Asking users to perform two tasks may have led users to rely on relatively simple strategies like \textit{cumulative probability} more than strategies which require more mental energy like \textit{frequency of draws changing order}. Conditions of high cognitive load seem to favor uncertainty visualizations like quantile dotplots over HOPs.

\textbf{Densities without means seem to support the best decision-making across levels of variance.} 
On a fixed axis scale, densities without means and quantile dotplots with means perform best at lower variance, while densities without means and intervals without means perform best at higher variance.
No visualization design we tested eliminated bias in decision-making at higher variance.
\textbf{Caveats:} The visualization design that leads to the least bias in decision-making depends on the variance of distributions relative to axis scale. %Since we only measure performance at two levels of variance and one axis scale, future work will be required to provide more granular design recommendations.
Future work should investigate bias in decision-making over a gradient of variances shown on a common scale, including charts with heterogeneous variances, as this would enable more exhaustive design recommendations. %about visualization design for decision aids as a function of the span of distributions relative to the axis. 

\textbf{Adding means leads to small biases in magnitude estimation and decision-making from distributional comparisons,} leading users to underestimate effect size and make less utility-optimal decisions in most in most cases we tested.
\textbf{Caveats:} Although the biasing effects of means are mostly negligible, our estimates of these biases are probably very conservative for two reasons: (1) added means were only highly salient in the HOPs condition; and (2) in the absence of added means, users already tend to rely on \textit{relative position}, a cue which the mean merely reinforces.
The effects of adding means on decision quality reverse at high versus low variance, so these biases may disappear for specific combinations of variance and axis scale.

\textbf{Users rely on distance between distributions as a proxy for effect size, so designers should note when this will be misleading and encourage more optimal strategies.}
Our quantitative analysis shows that adding means induces small but reliable biases in magnitude estimation, consistent with distance-based heuristics.
Our qualitative analysis of strategies verifies that the majority of users (357 of 442; 80.8\%) rely on distance between distributions or mean difference to judge effect size.
\textbf{Caveats:} Subtle design choices probably impact the tendency to rely on distance heuristics versus other strategies. For example, including a decision threshold annotation on our charts (Fig.~\ref{fig:distance_strategy}) may have encouraged users to judge effect size as \textit{cumulative probability}, rather than probability of superiority, contributing to underestimation bias. 

\subsection{Limitations}
% \alex{Subsume as caveat under 4. above}
% Since most users seem to not completely understand the statistical judgments they should be doing, our choice to include a decision threshold annotation on charts may have led users to judge effect size as cumulative probability rather than probability of superiority. 
% This may have contributed to underestimation bias. 

% \alex{Subsume as caveat under 3. above}
% The salience of added means may have varied across uncertainty visualizations, with only the means in the HOPs condition being highly salient.
% Because of this, our estimates of the impact of adding means averaging over visualization conditions are probably very conservative.

% We held axis limits constant across chart stimuli in order to study our hypothesis about users relying on distance as a proxy for effect size.
% In agreement with prior work~\cite{Witt2019,Hofman2020}, we think that the tendency to underestimate effect size may change depending on design choices about axis scale. 
% Our choice to hold axis limits constant does not impact comparisons between visualization designs in our study.
% \alex{This paragraph may not be necessary given the extensive discussion of axis scales in the caveats above.}

% Not investigating effects when distributions with heterogeneous variances are visualized in the same chart may limit the generalizability of our inferences. 
% Future work should study distributional comparisons with mixed levels of variance (e.g., in forest plots).

We only tested symmetrical distributions, and this may limit the generalizability of our inferences.
Although we speculate that chart users may rely on central tendency regardless of the family of a distribution, reasoning with multi-modal distributions in particular may involve different strategies not accounted for in the present study.

Because we rely on self-reported strategies in our qualitative analysis, our findings only reflect \textit{conscious} strategies.
This leaves out implicit or automatic information processing such as visual adaptation~\cite{Kale2019adaptation} and ensemble processing~\cite{szafir2016}, except in rare cases where users report trying to \textit{``roughly average''} predictions presented as HOPs.

Our choice to incentivize the decision-making task but not magnitude estimation may have contributed to the decoupling of performance on our two tasks.
We cannot disentangle this possible explanation from evidence corroborating Kahneman and Tversky's~\cite{Kahneman1979} distinction between perceived probabilities and decision weights (see Section 4.4).

We control the incentives for our decision task rather than manipulating them, in part because it is not feasible to test dramatically different incentives on Mechanical Turk. 
As such the risk preferences that we measure as PSEs are representative of users optimizing small monetary bonuses, and they may not capture how people respond to visualized data in crisis situations when lives, careers, or millions of dollars are at stake.
However, by devising a task that is representative of a broad class of decision problems (see Section 3.2), we make our results as broadly applicable as possible.
We speculate that the \textit{relative impacts} of visualization designs on risk preferences should generalize to decision problems with similar utility functions.

\subsection{Satisficing and Heterogeneity}
The visual reasoning strategies that chart users rely on when making judgments from uncertainty visualizations may not be what visualization designers expect. 
We present evidence that, in the absence of training, users satisfice by using suboptimal heuristics to decode the signal from a chart.
We also find that not all users rely on the same strategies and that many users switch between strategies.
Satisficing and heterogeneity in heuristics make it difficult both to anticipate how people will read charts and to study the impact of design choices.
Conventionally, visualization research has characterized visualization effectiveness by ranking visualization designs based on the performance of the average user (e.g.,~\cite{Cleveland1984}).
However, in cases like the present study where users are heterogeneous in their strategies, these \textit{averages may not account for the experience of very many users} and are probably an oversimplification.
Visualization researchers should be mindful of satisficing and heterogeneity in users' visual reasoning strategies, attempt to model these strategies, and try to design ways of training users to employ more optimal strategies.

\subsection{Toward Better Models of Visualization Effectiveness}
Because some users seem to adopt suboptimal strategies or switch between strategies when presented with an uncertainty visualization, models of visualization effectiveness which codify design knowledge and drive automated visualization recommendation and authoring systems should represent these strategies.
We envision a new class of behavioral models for visualization research which attempt to enumerate possible strategies, such as those we identify in our qualitative analysis, and learn how often users employ them to perform a specific task when presented with a particular visualization design.
Previous work~\cite{Jardine2020} demonstrates a related approach by calculating expected responses based on a set of alternative perceptual proxies for visual comparison and comparing these expectations to users' actual responses.
Like the present study, this work describes the correspondence between expected patterns and user behavior. 
Instead, we propose incorporating functions representing predefined strategies into predictive models which estimate the proportion of users employing a given strategy.

In a pilot study, we attempted to build such a model: a Bayesian mixture model of alternative strategy functions.
However, because multiple strategies predict similar patterns of responses, we were not able to fit the model due to problems with identifiability.
This suggests that the kind of model we propose will only be feasible if we design experiments such that alternative strategies predict sufficiently different patterns of responses.
The approach of looking at the agreement between proxies and human behavior~\cite{Jardine2020} suffers the same limitation, but there is no analogous mechanism to identifiability in Bayesian models to act as a fail-safe against unwarranted inferences. 
Future work should continue pursuing this kind of strategy-aware behavioral modeling.

We want to emphasize that the proposed modeling approach is not strictly quantitative, as the definition of strategy functions requires a descriptive understanding of users' visual reasoning.
As such this approach offers a way to formalize the insights of qualitative analysis and represent the gamut of possible user behaviors inside of visualization recommendation and authoring systems.

\section{Conclusion}
We contribute findings from a mixed design experiment on Mechanical Turk investigating how visualization design impacts judgments and decisions from effect size.
Our results suggest that visualization designs which support the least biased estimation of effect size do not necessarily support the best decision-making.
% , among the visualizations we tested, quantile dotplots support the least biased estimation of effect size, whereas densities and intervals support the most utility-optimal decision-making based on effect size.
We discuss how a user’s sense of the signal in a chart may not necessarily be identical when they use the same information for different tasks.
We also find that adding means to uncertainty visualizations induces small but reliable biases consistent with users relying on visual distance between distributions as a proxy for effect size.
In a qualitative analysis of users' visual reasoning strategies, we find that many users switch strategies and do not employ an optimal strategy when one exists.
We discuss ways that canonical characterizations of graphical perception in terms of average performance gloss over possible heterogeneity in user behavior, and we propose opportunities to build strategy-aware models of visualization effectiveness which could be used to formalize design knowledge in visualization recommendation and authoring systems beyond context-agnostic rankings of chart types.